\documentstyle{article}

\newcommand{\tr}{\mbox{\rm tr}}

\newcommand{\be}{\begin{equation}}
\newcommand{\ee}{\end{equation}}
\newcommand{\ben}{\begin{eqnarray}\displaystyle}
\newcommand{\een}{\end{eqnarray}}
\newcommand{\refb}[1]{(\ref{#1})}
\newcommand{\p}{\partial}

\begin{document}

\thispagestyle{empty}

\begin{flushright}
hep-th/0106231\\
AEI-2001-067\\
MRI-P-010605
\end{flushright}

\vskip 3.5cm

\begin{center}
{\Large \bf Normalization of the Boundary}\\
\medskip
{\Large \bf  Superstring Field Theory Action}

\vspace*{6.0ex}

{\large \rm Debashis Ghoshal\footnote{{\tt ghoshal@mri.ernet.in}}}

\vspace*{1.5ex}

{\large \it Harish-Chandra Research Institute\footnote{formerly 
{\em Mehta Reasearch Institute of Mathematics \&\ 
Mathematical Physics.}}}\\
{\large \it  Chhatnag Road, Jhusi}\\
{\large\it Allahabad 211 019, India}

\vspace*{5.5ex}

{\bf Abstract}

\begin{quote}
The normalization of the boundary superstring field theory action
is determined by computing the near on-shell amplitude involving 
three gauge fields. 
\end{quote}

{\em Contribution to the proceedings of Strings 2001, Mumbai, India.}
\end{center}

\newpage


In the recent past there has been some progress in understanding the
behaviour of the tachyonic scalar field in open string theory. In the
bosonic theory, this scalar is the lowest mode on the 
D-$p$-branes. In superstring it appears on the non-BPS
D-$p$-branes as well as on non-supersymmetric D$p$-D$\bar p$
systems. Sen has
conjectured\cite{9902105} that there is a (local) minimum of the tachyon
potential which describes the closed string vacuum without any D-brane. 
At this minimum the negative contribution from the tachyon potential 
exactly cancels the tension of the D-brane(s). Further, a codimension 
one lump solution in the bosonic case describes a D-$(p-1)$-brane,
while in the superstring case, a kink solution represents a BPS 
D-$(p-1)$-brane (respectively a non-BPS D-($p-1$)-brane) in the same 
theory. This picture generalizes to solitons of higher codimension. 

While support for these conjectures come from the analysis of the 
worldsheet theory\cite{WORLD,9902105}, various toy models of tachyon 
condensation\cite{padic,TOY}, as well as noncommutative limit 
of the effective field theory of the tachyon\cite{NONCOM}, the
natural setting to prove these is off-shell string field theory
(SFT).

The most well understood SFT is the covariant open bosonic string
field theory\cite{SFT}. There has been numerical investigation in 
the level truncation scheme of this theory\cite{SFTAN} that provides
very good evidence in the favour of the conjectures. Similar analysis 
hass also been done for superstrings in a covariant open 
string field theory proposed by Berkovits\cite{BeSSFT}.

There is also a formulation of bosonic open string field theory 
based on the boundary perturbation of the worldsheet theory on the 
disc\cite{9208027,9210065}. It has been 
shown\cite{0009103,0009148} that it is possible to prove the 
conjectures for the bosonic open string in the framework of this so 
called boundary string field theory (B-SFT). A crucial ingredient 
in the exact verification of this idea is the determination of the 
normalization of the B-SFT action. This was computed in 
Ref.\cite{GSnorm}. 

Soon afterwards, (inspired by earlier works\cite{tseyt}), a proposal 
for the boundary superstring field theory (B-SSFT) was made and the 
behaviour of the tachyon potential analyzed\cite{B-SSFT}. 
More recently there have been attempts\cite{Ma-NiPr} to derive this 
action along the lines of Refs.\cite{9208027}. In this note we shall 
compute the normalization of this B-SSFT.

\bigskip

In order to define boundary superstring field theory, we take the 
worldsheet to be a disc of unit radius with a flat metric on it.
A field configuration of the string 
is now associated with a boundary operator of ghost number +1 
picture number
$-1$ in the worldsheet field theory of matter and ghost system.  
For definiteness, let us consider the boundary superstring 
theory of a non-BPS D-$p$-brane, and restrict to operators of the 
form ${\cal O}=c e^{-\phi}{\cal V}$, where $c$ is the ghost field, $\phi$ 
comes from the `bosonization' of the $\beta\gamma$ superghost system,
($\beta = (\partial\xi)e^{-\phi},\;
\gamma = \eta e^\phi$),
and ${\cal V}=\sum_\alpha\lambda^\alpha{\cal V}_\alpha$ is a 
boundary operator in 
the matter theory. We assume that the ghosts and 
matter are decoupled, and that $\{{\cal V}_\alpha\}$ denotes a complete
set of boundary vertex operators. The theory has superconformal
invariance which is in general broken on the boundary by boundary 
perturbations. If we use the Hilbert space of the SCFT in the absence
of boundary perturbation, a string field configuration 
associated with the operator ${\cal V}={\cal P}$, (${\cal P}$ a 
superconformal primary 
operator of the matter sector), is described by the worldsheet action 
\be \label{e1}
{\cal I}\; =\;
{\cal I}_{Bulk} - \int {d\tau\over2\pi}\int d\theta\,
{\cal P}\;
=\; {\cal I}_{Bulk} - \int {d\tau\over2\pi}
\left(G^{(m)}_{-1/2}{\cal P}\right) ,  
\ee
where the angle $0\le\tau\le 2\pi$ parameterizes the boundary of 
the disc, $\theta$ its fermionic superpartner, and
${\cal I}_{Bulk}$ denotes the bulk worldsheet action corresponding to
the closed string background.
We shall consider a trivial (closed string) background in flat space, 
therefore ${\cal I}_{Bulk}$ describes the CFT of ten free superfields 
($X^\mu+\theta\psi^\mu)$ and the $(b,c)$ and $(\beta,\gamma)$ 
ghost fields.

The B-SSFT action ${\cal S}_{B}(\lambda^\alpha)$ is
determined from the equation:
\be \label{e3}
{\delta{\cal S}_{B}\over \delta \lambda^\alpha} = -\,{K\over2} 
\int {d\tau\over2\pi} \int {d\tau'\over 2\pi} \left\langle 
c e^{-\phi}{\cal V}_\alpha(\tau) \{Q_B,c e^{-\phi}{\cal V}(\tau')\}
\right\rangle_{\cal V}\, ,
\ee
where $\langle\,\cdots\,\rangle_{\cal V}$ denotes correlation
function in the worldsheet field theory described by the action
\refb{e1}; $Q_B$ is the BRST charge defined from the bulk theory
\be\label{BRST}
Q_B = \oint {dz\over2\pi i}\left\{c\left(T^{(m)} + T^{(\xi\eta)} + 
T^{(\phi)}\right) + 
c(\partial c)b + \eta e^\phi G^{(m)} - \eta(\partial\eta)
e^{2\phi}b\right\},
\ee  
and $K$ is a normalization constant we would like to determine. 

Let us review the tachyon potential derived in 
Refs.\cite{B-SSFT,tseyt}.
Consider a tachyon configuration on a single non-BPS D-$p$-brane
\be\label{tachycon}
{\cal V} = {\cal T}(X+\theta\psi)\Gamma + \Gamma D\Gamma,
\ee
where $\Gamma=\mu + \theta F$ is the superfield for the so called
boundary fermion\cite{WitK,B-SSFT}, that is, $\mu$ is the boundary 
fermion and $F$, an auxiliary field, is its partner; and 
$D=\partial_\theta + \theta\partial_\tau$. 
In particular the linear tachyon
${\cal T} = a + u (X^1+\theta\psi^1)$
was analysed in Refs.\cite{B-SSFT}. With this perturbation
the worldsheet theory remains free and the action can be computed 
{\em exactly}. More precisely, for the constant mode of the tachyon
the action is\footnote{We use the convention that when the
boundary perturbation ${\cal V}=0$, the partition function of the matter 
superconformal theory on the unit disk is equal to the volume 
$\Omega_d$ of the $d=(p+1)$ dimensional world-voulme of the 
D-$p$-brane.} 
\be\label{tach-pot}
{\cal S}_B(a) = -\, K\,\Omega_d\;e^{-a^2/4} .
\ee
The tachyon potential has an unstable extremum at $a=0$
corresponding to the D-$p$-brane. 
There is another extremum at $a=\infty$, which we shall
call the {\em vacuum} solution. The difference in energy density 
between the extremum $a=0$ and the vacuum is
$K$. Thus consistency with the conjectured behaviour would
require that $K$ equals $T_p$, the tension of the non-BPS
D-$p$-brane.

In the following we shall determine the normalization $K$. An 
analogous exercise was carried out in Ref.\cite{GSnorm} for the 
bosonic case\footnote{This was the subject of the talk in {\em
Strings 2001}.}. This
was done by computing the on-shell three tachyon amplitude from the 
cubic open string field theory\cite{SFT}, and comparing it with the 
same amplitude computed in the B-SFT\cite{9208027}. 
An exact repetition of this unfortunately does not 
work in the superstring case, where the tachyon lives 
in the GSO odd sector. Therefore the three tachyon amplitude vanishes, 
and the first non-trivial amplitude involves four tachyons. This, however,
receives contribution from (infinitely many) three-point vertices. 
Nevertheless the answer is computable in Berkovits' covariant superstring 
field theory\cite{BeEc}\footnote{It is not known
how to treat the Ramond sector in this formalism, but thankfully that 
will not be needed for our purpose.}, and of course in the 
first quantized theory since we are interested in the on-shell 
amplitude. But the same computation is difficult, if not
impossible, with our present understanding of B-SSFT.

Instead we shall consider the GSO even vector field, for definiteness,
on $N$ non-BPS D-$p$-branes of the appropriate type II superstring
theory. This has non-zero two and three point functions, which we 
shall compute in covariant and boundary superstring field theories in 
the remainder of this note. Since our treatment closely follows 
Ref.\cite{GSnorm}, we shall be brief. 

\bigskip

Let us consider the a near on-shell gauge field configuration. The
matter part of the corresponding vertex operator is 
\be\label{boungauge}
{\cal P} = \int{d^{d}k\over(2\pi)^{d}} \, 
{\cal A}_\mu^a(k) t^a\, \psi^\mu e^{ik\cdot X}\, , 
\ee
with ${\cal A}_\mu^a(k)$ supported over near on-shell momenta 
$k^2\simeq 0$. The Chan-Paton matrices $t^a$ are generators of the 
U($N$) Lie algebra. {}From eqn.\refb{e1}, we see that the boundary 
perturbation is:
\be \label{e11}
\int {d\tau\over 2\pi}\int d\theta\, {\cal P} =
\int {d\tau\over 2\pi} \int{d^{d}k\over(2\pi)^{d}} \, 
{\cal A}_\mu^a(k) t^a\, \left(i\p_\tau X^\mu + k_\sigma
\psi^\sigma\psi^\mu\right) e^{ik\cdot X}(\tau)\, ,
\ee
The configuration where ${\cal A}_\mu(k)\propto k_\mu$, yields a
perturbation by a total derivative term. We ignore these
in the following as we assume the covariant gauge condition
$k\cdot{\cal A}(k)=0$.

To calculate the quadratic term in B-SSFT, it is sufficient to
evaluate the correlator in eqn.(\ref{e3}) in the absence of any
boundary perturbation. We also use the fact that only the picture
number zero piece of the BRST operator contributes, therefore
\be \label{e12}
\{Q_B, c e^{-\phi}\psi^\mu e^{ik\cdot X}(\tau)\} = k^2 
(\p_\tau c) c e^{-\phi}\psi^\mu e^{ik\cdot X}(\tau) +\cdots\, .
\ee
To leading order in $k^2$, we find the near on-shell quadratic term:
\be \label{e13}
{\cal S}^{(2)}_{B} = {K\over4} \int{d^{d}k\over(2\pi)^{d}}\,
(-k^2){\cal A}_\mu^a(k){\cal A}_\nu^b(-k)\eta^{\mu\nu}\tr(t^at^b)
+\cdots\, ,
\ee 
for the gauge fields in B-SSFT.

\bigskip

Next we would like to evaluate the on-shell coupling of three gauge
fields. As in Ref.\cite{GSnorm}, a direct determination of this coupling 
is difficult due to the problem with ultraviolet divergences in the
correlator in eqn.(\ref{e3}). Fortunately a way around was found there.
We use the fact that the equations of motion derived from the 
B-SSFT action ${\cal S}_{B}$ must be proportional to the 
$\beta$-functions of the boundary theory described by the action 
\refb{e1}. At this point we need the following result from perturbed 
superconformal field theories\cite{BETA}. If an SCFT is perturbed by a
nearly marginal primary operator ${\cal V}_\alpha$ of dimension 
$h_\alpha\simeq 1/2$, the $\beta$-function associated with the 
coupling $\lambda^\alpha$, to second order\footnote{To this order, the
$\beta$-function is free of regularization ambiguity in taming 
ultraviolet divergence.} in $\lambda$, is given by
\be \label{e16}
{\delta{\cal S}_B\over\delta\lambda_\alpha}\;\propto\;
\beta^\alpha(\lambda)\; \propto\; \left(h_\alpha - {1\over2}\right) 
\lambda^\alpha + {1\over 2\pi} C^\alpha_{\beta\gamma}
\lambda^\beta\lambda^\gamma\, ,
\ee
where $C^\alpha_{\beta\gamma}$ is determined from the operator
product expansion of the unperturbed SCFT 
\be\label{e15}
{\cal V}_\beta(\tau)\left(G^{(m)}_{-1/2}{\cal V}_\gamma\right)(\tau') \simeq 
{C^\alpha_{\beta\gamma}\over
\left[2\sin\left({\tau-\tau'\over2}\right)\right]^{-h_\alpha+
h_\beta+h_\gamma+1/2}}\;{\cal V}_\alpha(\tau').
\ee
(The factor of $1/2\pi$ in front of the  second term is due to
the same normalization factor in the boundary perturbation 
\refb{e1}.)
Our gauge field vertex operators \refb{boungauge} have 
conformal weights $h(k)=k^2+{1\over2}\simeq {1\over2}$, and the
necessary OPE is
\begin{eqnarray} \label{e17}
\psi^\nu e^{ik_2\cdot X}(\tau)\!\!\!\!\! &{}&\!\!\!\!\! 
\left(i\p_{\tau'} X^\rho + k_{3\sigma}
\psi^\sigma\psi^\rho\right) e^{ik_3\cdot X}(\tau')\\ 
\simeq &{}&
{(2\pi)^d\delta^{(d)}(k-k_2-k_3)\over 
2\sin\left({\tau-\tau'\over2}\right)}\,
\left(-\delta_\mu^\nu k_2^\rho - \eta^{\nu\rho}k_{3\mu} + 
\delta^\rho_\mu k_3^\nu\right)\,\psi^\mu e^{ik\cdot X}(\tau')
\, .\nonumber
\end{eqnarray}
The equation of motion for the (near on-shell) gauge field 
\refb{boungauge} is therefore
\begin{eqnarray} \label{e18}
&&k^2\eta^{\mu\nu} {\cal A}_\nu^b(k)t^b + {1\over 2\pi} 
\int{d^{d}k_2\over(2\pi)^{d}}\int{d^{d}k_3\over(2\pi)^{d}} 
(2\pi)^{d}\delta^{(d)}(k-k_2-k_3)\nonumber\\
&& \qquad\qquad\times\, 
(-\eta^{\mu\nu}k^\rho + \eta^{\nu\rho}k_2^\mu + \eta^{\rho\mu}k_3^\nu)\,
{\cal A}_\nu^b(k_2)t^b\, {\cal A}_\rho^c(k_3)t^c = 0\, .
\end{eqnarray}
In the above we have used the momentum constraint and the
gauge condition $k\cdot{\cal A}(k)=0$, and also relabelled dummy
variables in the second term.

We integrate \refb{e18} to obtain the (gauge fixed) 
action upto cubic terms in gauge fields. The result is
\begin{eqnarray}\label{b-action}
{\cal S}_{B} &\simeq& {K\over2} \Big[ {1\over2}
\int {d^{d}k\over(2\pi)^{d}}\, (-k^2)\, {\cal A}_\mu^a(k)
{\cal A}_\nu^b(-k)\eta^{\mu\nu}\tr(t^at^b) \nonumber \\
&&\!\!\!\! -\,{1\over 6\pi} \int {d^{d}k_1\over(2\pi)^{d}} 
\int{d^{d}k_2\over(2\pi)^{d}} \int{d^{d}k_3\over(2\pi)^{d}} \, 
(2\pi)^{d}\delta^{(d)}(\sum k_i)\,\tr(t^at^bt^c)\nonumber\\
&&\qquad\times\; \left(\eta^{\mu\nu}k_1^\rho +
\eta^{\nu\rho}k_2^\mu+\eta^{\rho\mu}k_3^\nu\right)
{\cal A}_\mu^a(k_1){\cal A}_\nu^b(k_2) 
{\cal A}_\rho^c(k_3)\Big]\, ,
\een
to leading order in $k^2$.
The overall normalization in \refb{b-action} has been fixed by 
using eqn.\refb{e13}. The above action may now be compared this 
with results from covariant open SSFT. 

\bigskip

The relevant part of the action of the covariant open SSFT 
is\cite{BeSSFT}
\begin{eqnarray} \label{e7}
S_{Be}&=&
2\pi^2T_p\,\Big[{1\over2!}\left\langle\left\langle(Q_B\Phi)(\eta_0\Phi)
\right\rangle\right\rangle\; + \nonumber\\ 
&{}&\quad{1\over3!}\Big(\left\langle\left\langle
(Q_B\Phi)\Phi(\eta_0\Phi)\right\rangle\right\rangle -
\left\langle\left\langle(Q_B\Phi)(\eta_0\Phi)\Phi
\right\rangle\right\rangle\Big)\,+\,\cdots\Big] ,
\end{eqnarray}
where $|\Phi\rangle$ is the string field represented by a ghost 
number zero picture number zero state in the Hilbert space of the first 
quantised theory. The correlators $\langle\langle\cdot\rangle\rangle$
are defined through known conformal maps reviewed in Ref.\cite{BeSSFT}, 
where the normalization factor $2\pi^2 T_p$ was derived.

We now consider a gauge string field configuration of the form
\be \label{e9}
|\Phi\rangle = \int{d^{d}k\over(2\pi)^{d}} \, 
A_\mu^a(k)t^a\,\xi_0 c_1 e^{-\phi}\psi^\mu_{-1/2}|k\rangle\, ,
\ee
with wavefunction $A_\mu^a(k)$ supported over near on-shell momenta 
$k^2\simeq 0$. We use the string field configuration \refb{e9} in 
\refb{e7}, and keep only the leading order terms in $k^2$ to
arrive at the action:
\ben \label{e10}
S_{Be} &\simeq& 2\pi^2 T_p \Big[ {1\over2}
\int {d^{d}k\over(2\pi)^{d}}\, (-k^2)\, A_\mu^a(k)
A_\nu^b(-k)\eta^{\mu\nu}\tr(t^at^b) \nonumber \\
&&\!\!\!\! -\,{1\over 3} \int {d^{d}k_1\over(2\pi)^{d}} 
\int{d^{d}k_2\over(2\pi)^{d}} \int{d^{d}k_3\over(2\pi)^{d}} \, 
(2\pi)^{d}\delta^{(d)}(\sum k_i)\,\tr(t^at^bt^c)\nonumber\\
&&\qquad\times\; \left(\eta^{\mu\nu}k_1^\rho +
\eta^{\nu\rho}k_2^\mu+\eta^{\rho\mu}k_3^\nu\right)
A_\mu^a(k_1)A_\nu^b(k_2) 
A_\rho^c(k_3)\Big]\, .
\een
In the above we have used the Siegel gauge, which for the
gauge field means that $k^\mu A_\mu^a(k) = 0$.

%
%

\bigskip

A comparison between the quadtratic terms in the actions 
\refb{b-action} and \refb{e10} relates the gauge string fields 
in the two formalisms:
\be\label{fieldrel}
{\cal A}_\mu^a(k) = 2\pi\,\sqrt{{T_p\over K}}\, A_\mu^a(k) + \cdots ,
\ee
to leading order. The cubic interaction term now determines the
required normalization $K=T_p$. 

\bigskip

\noindent{\bf Acknowledgement:} It is a pleasure to thank Ashoke Sen
for collaboration on Ref.\cite{GSnorm} and for valuable discussions
and suggestions. I am grateful to Stefan Theisen and the
Albert-Einstein-Institut, Golm, Germany, for their kind hospitality
during this work.

\end{document}